%% file: transportmovinglattice.tex
\documentclass[12pt]{iopart}
\usepackage{iopams}
\usepackage[dvips]{graphicx}

\begin{document}

\newcommand{\bra}[1]{\langle #1|}
\newcommand{\ket}[1]{|#1\rangle}
\newcommand{\braket}[2]{\langle #1|#2\rangle}

\title{Long Distance Transport of Ultracold Atoms using a 1D optical lattice}

\author{Stefan Schmid, Gregor Thalhammer, Klaus
Winkler,\\ Florian Lang, and Johannes Hecker Denschlag}
\address{Institut f\"ur Experimentalphysik, Universit\"at Innsbruck,
 Technikerstra{\ss}e 25,\\ 6020 Innsbruck,
Austria}

\date{\today}

\begin{abstract}
We study the horizontal transport of ultracold atoms over
macroscopic distances of up to 20$\,$cm with a moving 1D optical
lattice. By using an optical Bessel beam to form the optical
lattice, we can achieve nearly homogeneous trapping conditions over
the full transport length, which is crucial in order to hold the
atoms against gravity for such a wide range. Fast transport
velocities of up to 6$\,$m/s (corresponding to about 1100 photon
recoils) and accelerations of up to 2600$\,$m/s$^2$ are reached.
Even at high velocities the momentum of the atoms is precisely
defined with an uncertainty of less than one photon recoil. This
allows for construction of an atom catapult with high kinetic energy
resolution, which might have applications in novel collision
experiments.
\end{abstract}

\tableofcontents

\newpage

\section{Introduction}

Fast, large-distance transport of Bose-Einstein condensates (BEC)
from their place of production to other locations is of central
interest in the field of ultracold atoms. It allows for exposure of
BECs to all different kinds of environments, spawning progress in
BEC manipulation and probing.

Transport of cold atoms has already been explored in various
approaches using magnetic and optical fields. Magnetic fields have
been used to shift atoms, e.g. on atom chips  (for a review see
\cite{Mag02}) and to move laser cooled clouds of atoms over
macroscopic distances of tens of centimeters, e.g.
\cite{Gre01,Lew02}. By changing the position of an optical dipole
trap, a BEC has been transferred over distances of about 40$\,$cm
within several seconds \cite{Gus02}. This approach consisted of
mechanically relocating the focussing lens of the dipole trap with a
large translation stage. A moving optical lattice offers another
interesting possibility to transport ultracold atoms. Acceleration
of atoms with lattices is intimately connected to the techniques of
Raman transitions \cite{Kas91}, STIRAP \cite{Ber98,Pot01} and the
phenomenon of Bloch oscillations \cite{Dah96,Mor01}; (for a recent
review on atoms in optical lattices see \cite{Mor06}). Acceleration
with optical lattices allows for precise momentum transfer in
multiples of 2 photon recoils to the atoms. Transport of single,
laser cooled atoms in a deep optical lattice over short distances of
 several mm has been reported in \cite{Kuh01}. Coherent transport
of atoms over several lattice sites has been described in
\cite{Man03}. Even beyond the field of ultracold atoms, applications
of optical lattices for transport are of interest, e.g. to relocate
sub-micron sized polystyrene spheres immersed in heavy water
\cite{Ciz05}.
\\Here we experimentally investigate transporting BECs
 and ultracold thermal samples with an optical lattice over
 macroscopic distances of tens of cm. Our method features the
 combination of the following important characteristics. The transport of the atoms
is in the quantum regime, where all atoms are in the vibrational
ground state of the lattice. With our setup, mechanical
 noise is avoided and we achieve precise positioning
(below the imaging resolution of 1$\,\mu$m). We demonstrate high
 transport velocities of up to 6$\,$m/s, which are accurately controlled on the quantum
 level. The velocity spread of the atoms is
 not more than 2$\,$mm/s, corresponding to 1/3 of a
photon recoil.

\begin{figure}
\centering
\includegraphics{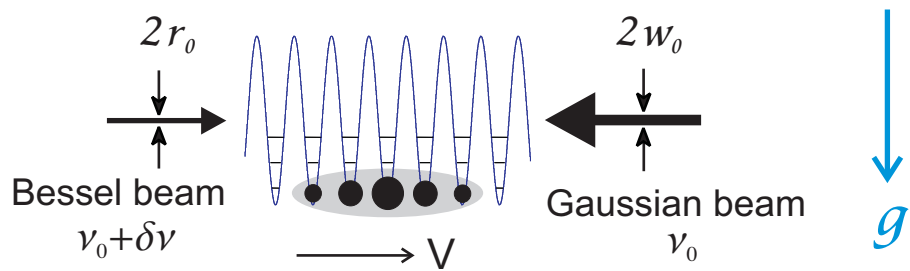}
\caption{\textsc{Scheme for atom transport.} Two counterpropagating
 laser beams form a standing wave
dipole trap. A BEC is loaded adiabatically into the vibrational
ground state of this 1D optical lattice. A relative frequency
detuning $\Delta \nu$ between the two laser beams results in a
lattice motion at a velocity $v=\Delta\nu\cdot\lambda/2$ which drags
along the trapped atoms. We chose the counterpropagating laser beams
to consist of a Gaussian beam with diameter 2$w_{0}$ and a Bessel
beam with a central spot diameter of 2$r_{0}$. The (in principle)
diffraction-free propagation of the Bessel beam leads to tight
radial confinement of the atoms over long distances, which supports
the atoms against gravity during horizontal transport. }
\label{fig:skizze1}
\end{figure}

\section{Basic Principle of Transport}
Horizontal transport of atoms over larger distances holds two
challenges: how to move the atoms and how to support them against
gravity. Our approach here is to use a special 1D optical lattice
trap, which is formed by a Bessel laser beam and a
counterpropagating Gaussian beam. The lattice part of the trap moves
the atoms axially, whereas the Bessel beam leads to radial
confinement holding the atoms against gravity.

In brief, lattice transport works like this. We first load the atoms
into a 1D optical lattice, which in general is a standing wave
interference pattern of two counter propagating laser beams far
red-detuned from the atomic resonance line (see Figure 1).
Afterwards the optical lattice is carefully moved, ``dragging''
along the atoms. Upon arrival at the destination, the atoms are
released from the lattice.

The lattice motion is induced by dynamically changing the relative
frequency detuning $\Delta \nu$ of the two laser beams, which
corresponds to a lattice velocity
\begin{equation}
v = \frac{\lambda}{2}\Delta\nu \label{vDelta},
\label{velocity-detuning}
\end{equation}
where $\lambda$ is the laser wavelength of the lattice.

In comparison to the classical notion of simply ``dragging'' along
the atoms in the lattice, atom transport is more subtle on the
quantum level. Here only momenta in multiples of two photon recoil
momenta, $2 \hbar k = 4 \pi \hbar / \lambda $ can be transferred to
the atoms. This quantized momentum transfer can be understood in
several ways, e.g. based on stimulated Raman transitions or based on
the concept of Bloch-like oscillations in lattice potentials. For a
more thorough discussion in this context the reader is referred to
\cite{Den02}.

In order to prevent the atoms from falling in the gravitational
field, the lattice has to act as an optical dipole trap in the
radial direction. It turns out that for radial trapping, optical
lattices formed by Bessel beams have a clear advantage over Gaussian
beam lattices. To make this point clear, we now show, that a
standard optical lattice based on Gaussian beams is not well suited
for long distance transports on the order of 50$\,$cm. During
transport we require the maximum radial confining force
$F_{\textrm{\scriptsize max}}$ to be larger than gravity $m g$,
where $m$ is the atomic mass and $g\approx$ 9.81$\,$m/s$^2$ is the
acceleration due to gravity. For a Gaussian beam this is
\begin{eqnarray}
 F_{\textmd{\scriptsize max}} =
 \frac{3}{4\pi^{3}\sqrt{\textrm{e}}}\,\frac{\lambda^{3}}{c}\,\frac{\Gamma}{\Delta}\,\frac{P_{0}}{w(z)^{3}} > m g
\end{eqnarray}
where $\Gamma$ is the natural linewidth of the relevant atomic
transition, $\Delta$ the detuning from this transition, $w(z)$ the
beam radius and $P_{0}$ the total power of the beam. The strong
dependence on the beam radius $w(z)$ suggests, that $w(z) =
w_{0}\sqrt{1+(z/z_{\textrm{\scriptsize R}})^{2}}$ should not vary
too much over the transport distance. If we thus require the
Rayleigh range $z_{\textrm{\scriptsize R}}=\pi w_{0}^{2}/\lambda$ to
equal the distance of 25$\,$cm, the waist has to be $w_{0} \approx$
260$\,\mu$m. For a lattice beam wavelength of e.g. $\lambda = 830
\,$nm, the detuning from the D-lines of $^{87}$Rb is $\Delta \approx
2\pi\times 130\,$THz. To hold the atoms against gravity for all $z$,
where $|z|<z_{\textrm{\scriptsize R}}$, a total laser power of
$P_{0} \approx$ 10$\,$W is needed, which is difficult to produce. In
addition the spontaneous photon scattering rate
\begin{equation}
 \Gamma_{scatt} = \frac{3}{8\pi^{3}\hbar}\,\frac{\lambda^{3}}{c}\,\left(\frac{\Gamma}{\Delta}\right)^{2}\,\frac{P_{0}}{w(z)^{2}}\\
\end{equation}
would reach values on the order of $\Gamma_{\textrm{\scriptsize
scatt}} =$ 2$\,$s$^{-1}$. For typical transport times of 200$\,$ms
this means substantial heating and atomic losses.

A better choice for transport are zero order Bessel beams (Figure
2). They exhibit an intensity pattern which consists of an inner
intensity spot surrounded by concentric rings and which does {\em
not} change during propagation. In our experiments we have formed a
standing light wave by interfering a Bessel beam with a
counter-propagating Gaussian beam, giving rise to an optical lattice
which is radially modulated according to the Bessel beam
\footnote{In principle, one could also use a pure Bessel lattice
(produced by two counter-propagating Bessel beams) for transport.
This would improve radial confinement, however, alignment is more
involved.}. Atoms loaded into the tightly confined inner spot of the
Bessel beam can be held against gravity for moderate light
intensities, which minimizes the spontaneous photon scattering rate.
In comparison to the
 transport with a Gaussian beam, the scattering rate in a Bessel beam transport
  can be kept as low as 0.05$\,$s$^{-1}$ by using the beam
  parameters of our experiment.

\begin{figure}
\centering
\includegraphics{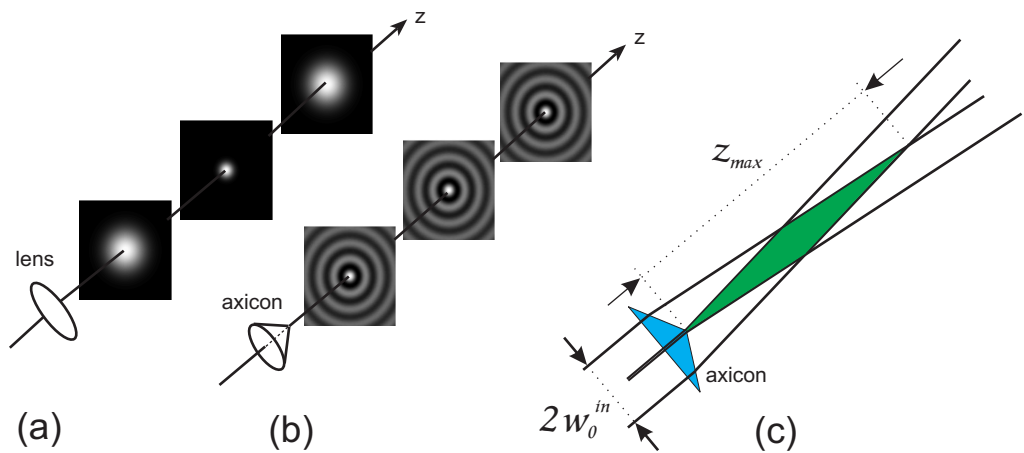}
\caption{ \textsc{Gaussian and Bessel beams.} \textbf{(a)} The
radial intensity distribution of a Gaussian beam changes as it
propagates. The smaller the waist $w_{0}$ of the beam, the higher
its divergence (for a given wavelength). \textbf{(b)} Bessel beam:
The radial distribution and in particular the radius of the central
spot $r_{0}$ do not change with $z$ (see equation
(\ref{besselradius})). \textbf{(c)}  Within a certain axial range
$z_{max}$ a Bessel-like beam can be produced by illuminating an
axicon lens with a collimated laser beam.}
\label{fig:gaussbesselskizze}
\end{figure}

\section{Bessel beams}

Bessel beams are a solution of the Helmholtz equation and were first
discussed and experimentally investigated about two decades ago
\cite{Dur87,Dur87a}.

In cylindrical coordinates the electric field distribution of a
Bessel beam of order $l$ is given by
\begin{eqnarray} E(r, \varphi, z)&=&E_0\ \ e^{i\beta z}\
e^{il\varphi}\ J_{l}(\alpha r) \ ,
\end{eqnarray}
where $J_{l}(\alpha r)$ is the Bessel function of the first kind
with integer order $l$. The beam is characterized by the parameters
$\alpha$ and $\beta$. In the following, we restrict the discussion
to order $l = 0$ which we have used in the experiment. By taking the
absolute square of this expression one gets the intensity
distribution given by
\begin{equation}
I(r,z) = I_{0}\; J_{0}^{2}(\alpha r) , \label{Ibessel}
\end{equation}
where $\alpha$ determines the radius $r_0$ of the central spot via
the first zero crossing of $J_{0}(\alpha r)$
\begin{eqnarray}
r_0 \approx {4.81 \over 2 \alpha}.
\end{eqnarray}
As pointed out before, $r_{0}$ and $I_{0}$ do not change with the
axial position $z$. Because of this axial independence the Bessel
beams are said to be ``diffraction-free".

Bessel-like beams were realized experimentally for the first time by
illuminating a circular slit \cite{Dur87a}. Since this method is
very inefficient, two other ways are common nowadays. To generate
Bessel beams of arbitrary order holographic elements, such as
phase-gratings, are used \cite{Nig97}. In our setup we use a zero
order Bessel beam, which can be produced efficiently by simply
illuminating an axicon (conical lens) with a collimated laser beam
\cite{Nig99}. How this comes about, can be understood by looking at
the Fourier transform of the Bessel field
\begin{eqnarray}
\widetilde{E}(k_{\bot},\varphi_{k},k_{z})&=&\int d^{3}r~
E(r,\varphi,z) e^{-ik_{\bot}r\cos(\varphi - \varphi_{k})}e^{-i
k_{z}z}\nonumber\\
&\propto& e^{il\varphi_{k}}~\delta(k_{z}-\beta)~ \delta
(k_{\bot}-\alpha). \label{equ:FTbessel}
\end{eqnarray}
  Thus a Bessel beam is a superposition of plane waves with
$(k_{\bot},k_{z})=(\alpha,\beta)$. The $\mathbf{k}$~-~vectors of the
plane waves all have the same magnitude $|\mathbf{k}|$ = $k$ =
$2\pi/\lambda$ = $\sqrt{\alpha^{2}+\beta^{2}}$ and they are forming
a cone with radius $k_{\bot}$ and height $k_{z}$. Using an axicon
with apex angle $\delta$ and index of refraction $n$, $\alpha$ and
$\beta$ are given by
\begin{equation}
\alpha = \frac{\pi(n-1)} {2 \lambda\tan\delta/2}
\label{besselradius}
\end{equation}
and
\begin{equation}
\beta = \sqrt{k^{2}-\alpha^{2}}.\label{besselbeta}
\end{equation}

These experimentally produced Bessel beams are not ideal in the
sense that their range $z_{\textrm{\scriptsize max}} = k
w_{0}^{\textrm{\scriptsize in}}/\alpha$ is limited by the finite
size (waist $w_{0}^{\textrm{\scriptsize in}}$) of the beam impinging
on the axicon lens (see Figure \ref{fig:gaussbesselskizze}(c)).
Also, the intensity of the Bessel beams might not be independent of
the axial coordinate $z$, as it is also determined by the radial
intensity distribution of the impinging beam (e.g. see Figure
\ref{fig:transport24a}(b)).

\section{Experimental setup}

We work with a $^{87}$Rb-BEC in the internal state $\ket{F=1,~
m_{F}=-1}$, initially held in a Ioffe-type magnetic trap with trap
frequencies of $2\pi\,\nu_{x,y,z}$ = $2\pi\,$(7$\,$Hz, 19$\,$Hz,
20$\,$Hz) \cite{Tha06,Tha05}. From the magnetic trap the condensate
is adiabatically loaded in about 100$\,$ms into the inner core of
the 1D optical lattice formed by a Bessel beam of central spot
radius $r_{0}=$ 36$\,\mu$m and a counter-propagating Gaussian beam
with a waist of $w_{0} = $ 85$\,\mu$m. About 70 lattice sites are
occupied with atoms in the vibrational ground state. The lattice
periodicity is 415$\,$nm, corresponding to the laser wavelength of
830$\,$nm. For our geometry (see below) the total power needed for
the Bessel beam to support the atoms against gravity is typically
200$\,$mW, since only a few percent ($\approx$ 10$\,$mW) of the
total power are stored in the central spot. For the Gaussian beam a
power of roughly 20$\,$mW is chosen, leading to an optical  trapping
potential at the center ($r=0$) of $U(z)=
-U_{0}+U_{\textrm{\scriptsize latt}}\sin^2(kz)$, where the lattice
depth (effective axial trap depth) is $U_{\textrm{\scriptsize
latt}}\approx$ 10$\,E_{\textrm{\scriptsize r}}$ and the total trap
depth $U_{0}\approx$ 11$\,E_{\textrm{\scriptsize r}}$. Here
$E_{\textrm{\scriptsize r}}=(\hbar k)^{2}/(2m)$ is the recoil
energy.

The corresponding trap frequencies are $\nu_{\bot}=
4.81\sqrt{U_{0}/(8m r_{0}^{2})}/(2\pi)$ = 97$\,$Hz in the radial
direction and $\nu_{z}=k\sqrt{2U_{\textrm{\scriptsize
latt}}/m}/(2\pi)$ = 21$\,$kHz in the axial direction. In order to
better analyze the transport properties, we mostly perform round
trips, where the atoms are first moved to a distance $D$ and then
back to their initial spot, which lies in the field of vision of our
CCD camera. Once back, the atoms are adiabatically reloaded into the
Ioffe-type magnetic trap. To obtain the resulting atomic momentum
distribution a standard absorption imaging picture is taken after
sudden release from the magnetic trap and typically 12$\,$ms of
time-of-flight.

The lattice beams for the optical lattice are derived from a
Ti:Sapphire-laser operating at 830$\,$nm. The light is split into
two beams, each of which is controlled in amplitude, phase and
frequency with an acousto-optical modulator (AOM). For both AOMs the
radio-frequency (RF) driver consists of a home-built 300$\,$MHz
programmable frequency generator, which gives us full control over
amplitude, frequency and phase of the radio-wave at any instant of
time. The frequency generator is based on an AD9854 digital
synthesizer chip from Analog Devices and a 8-bit micro-controller
ATmega162 from Atmel, on which the desired frequency ramps are
stored and from which they are sent to the AD9854 upon request.
After passing the AOMs, the two laser beams are mode-cleaned in
single-mode fibers and converted into collimated Gaussian beams. One
of the Gaussian beams passes the axicon lens (apex angle =
178$^{\circ}$, radius = 25.4$\,$mm, Del Mar Photonics) with a waist
of $w_0^{\textrm{\scriptsize in}} = 2\,$mm, producing the Bessel
beam. From there the beam propagates towards the condensate, which
-before transport- is located 5$\,$cm away.

\begin{figure}
\centering
\includegraphics{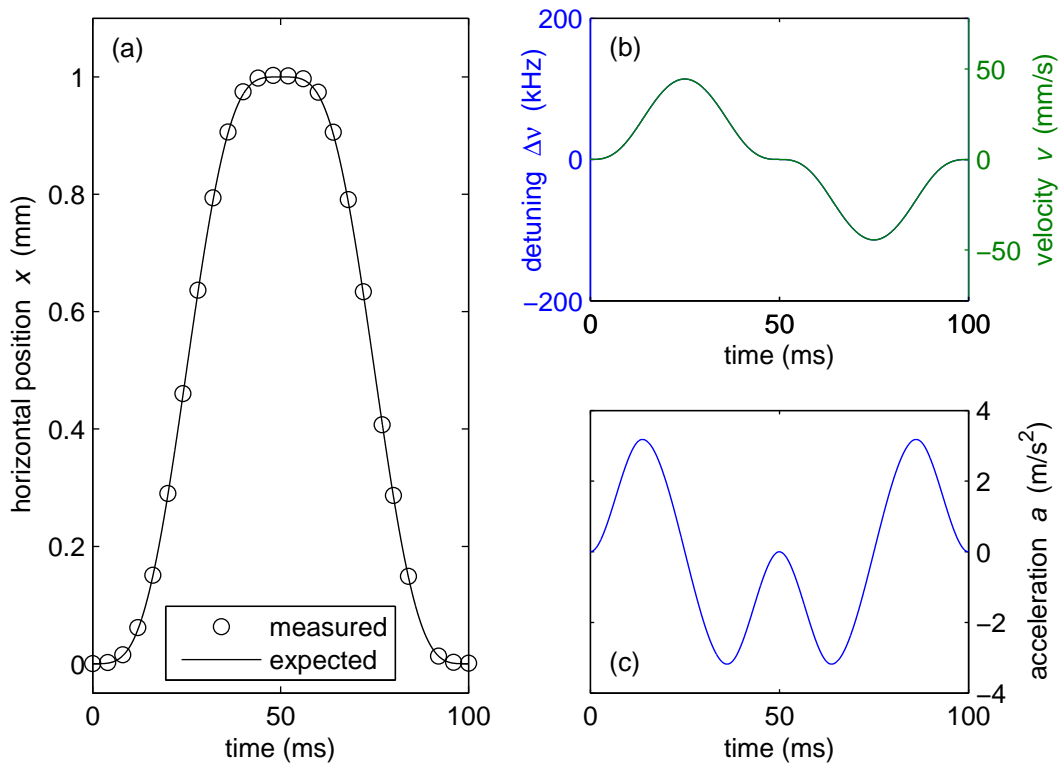}
\caption{\textbf{(a)} Position, \textbf{(b)} velocity and
\textbf{(c)} acceleration of the atomic cloud as a function of time
for a typical transport sequence, here a round-trip over a short
distance of 1mm. Piecewise defined cubic polynomials are used for
the acceleration ramp (see appendix for an analytical expression).
By integrating over time, velocity and position are obtained. The
frequency detuning $\Delta\nu$, which is used to program the RF
synthesizers, corresponds directly to the velocity $v$ via equation
(\ref{velocity-detuning}). The position ramp is compared with
in-situ measurements of the cloud's position
(circles).}\label{fig:transport43}
\end{figure}

\begin{figure}
\centering
\includegraphics{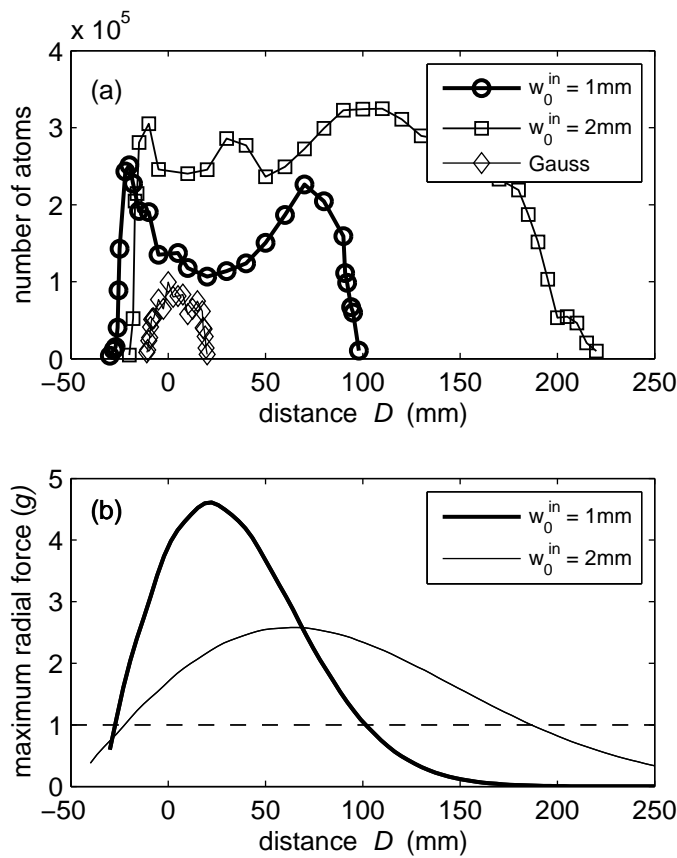}
\caption{ \textsc{Long distance transports.} \textbf{(a)} Shown is
the number of remaining atoms after a round-trip transport (see
Figure \ref{fig:transport43}) over various one-way distances $D$.
The first two data sets are obtained with two different Bessel beams
which are created by illuminating an axicon with a Gaussian beam
with a waist $w_{0}^{in}$ = 1$\,$mm and 2$\,$mm, respectively. The
transport time $T$ was kept constant at $T=$130$\,$ms and
$T=$280$\,$ms, respectively. The third data set (diamonds)
corresponds to a transport in a Gaussian beam lattice (see text).
 The calculated maximum radial trapping
force of the two Bessel beam lattice traps is shown in \textbf{(b)}
in units of $m g$, where $g \approx$ 9.81$\,$m/s$^2$ denotes the
gravitational acceleration. The variation of the trapping force with
distance is an imperfection of the Bessel beam and reflects its
creation from a Gaussian beam. When the maximum radial force drops
below 1$\,g$, gravity pulls the atoms out of the trap, as can be
clearly seen in (a). } \label{fig:transport24a}
\end{figure}

\section{Transport of ultracold atoms}

Figure \ref{fig:transport43} shows results of a first experiment,
where we have transported atoms over short distances of up to
1$\,$mm (round trip), so that they never leave the field of view of
the camera. The atoms move perpendicularly to the direction of
observation. In-situ images of the atomic cloud in the optical
lattice are taken at various times during transport and the center
of mass position of the cloud is determined. As is clear from Figure
\ref{fig:transport43}(a) we find very good agreement between the
expected and the measured position of the atoms. In Figure
\ref{fig:transport43}(b) and \ref{fig:transport43}(c), calculations
are shown for the corresponding velocity $v(t)$ and acceleration
$a(t)$ of the optical lattice, respectively. As discussed before
(see equation (\ref{vDelta})), the velocity $v$ of the lattice
translates directly into a relative detuning $\Delta \nu$ of the
laser beam, which we control via the AOMs. In order to suppress
unwanted heating and losses of atoms during transport, we have
chosen very smooth frequency ramps $\Delta\nu(t)$ such that the
acceleration is described by a cubic spline interpolation curve
which is continuously differentiable (details are given in the
appendix). In this way also the derivative of the acceleration
(commonly called the jerk) is kept small.

In the next set of experiments we have extended the atomic transport
to more macroscopic distances of up to 20$\,$cm (40$\,$cm round
trip), where we moved the atoms basically from one end of the vacuum
chamber to the other and back. However, the transport distance was
always limited by the finite range $z_{\textrm{\scriptsize max}}$ of
the Bessel beam (see Figure \ref{fig:gaussbesselskizze}(c) and
Figure \ref{fig:transport24a}). As shown in Figure
\ref{fig:transport24a}, the total number of atoms abruptly decreases
at the axial position, where the maximum radial force drops below
gravity. It is also clear from the figure, how the range of the
Bessel beam is increased by enlarging the waist
$w_0^{\textrm{\scriptsize in}}$ of the incoming Gaussian beam. Of
course, for a given total laser power, the maximum radial force
decreases as the Bessel beam diameter is increased. For the
transport distances of 12$\,$cm and 20$\,$cm the total power in the
Bessel beam was approximately 400$\,$mW. For comparison, we have
also transported atoms with a lattice formed by two
counter-propagating Gaussian beams (see Fig. 4 (a)). For this
transport both laser beams have a Rayleigh range of $z_{R}\approx
2\,$cm corresponding to a waist of 70 $\mu$m. The laser power of the
two beams was $\approx 130\,$mW and $\approx 35\,$mW, respectively.
We observe a sudden drop in atom number when the transport distance
exceeds the Rayleigh range. Using the scaling law given in equation
(2) it should be clear that transports of atoms over tens of cm with
a Gaussian lattice is hard to achieve.

Interestingly, the curve corresponding to the Bessel beam with waist
$w_{0}^{\textrm{\scriptsize in}}$ = 1$\,$mm in Figure
\ref{fig:transport24a}(a) exhibits a pronounced minimum in the
number of remaining atoms at a distance of about 3$\,$cm. The
position of this minimum coincides with the position, where the
lattice depth has a maximum (see Figure \ref{fig:transport24a}(b)).
This clearly indicates, that high light intensities adversely affect
atom lifetimes in the lattice. Although we have not studied in
detail the origin of the atomic losses in this work, they should
partially originate from spontaneous photon scattering and 3-body
recombination. In the deep lattice here (60$\,E_{\textrm{\scriptsize
r}}$) the calculated photon scattering rate is
$\Gamma_{\textrm{\scriptsize scatt}} =$ 0.4$\,$s$^{-1}$. The tight
lattice confinement leads to a high calculated atomic density of
$n_0 \approx 2 \ 10^{14}$ cm$^{-3}$. Adopting $L = 5.8\times
10^{-30}$cm$^6$/s as rate coefficient for the three body
recombination \cite{Sod99}, we expect a corresponding loss rate $L
n_{0}^{2}=$0.3$\,$s$^{-1}$.

\begin{figure}
\centering
\includegraphics{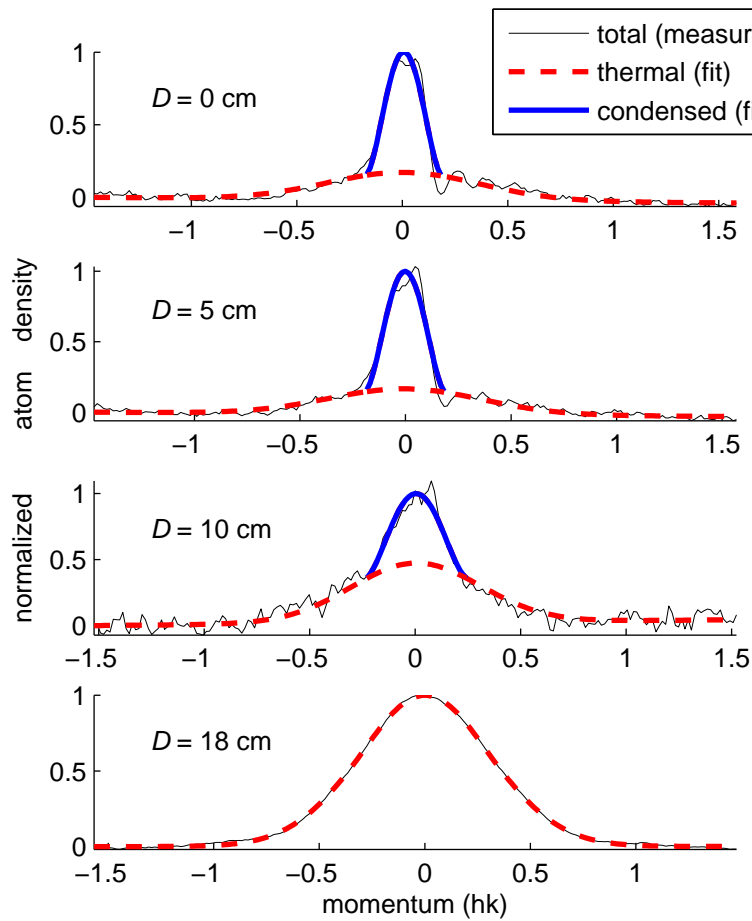}
\caption{\textsc{Transporting BEC.} Shown are the momentum
distributions (thin black lines) of the atoms  after a return-trip
transport over various one-way distances $D$. A bimodal distribution
(a blue parabolic distribution for the condensed fraction and a red
Gaussian distribution for the thermal fraction) is fit to the data.
For $D$ below 10cm a significant fraction of the atomic cloud is
still condensed. For $D = $18$\,$cm ($\approx$ the limit in our
experiments) only a thermal cloud remains, however, with a
temperature below the recoil limit (T $<$ 0.2$\,E_{r}/k_{B}$
$\approx$ 30$\,$nK).} \label{fig:becpics}
\end{figure}

\begin{figure}
\centering
\includegraphics{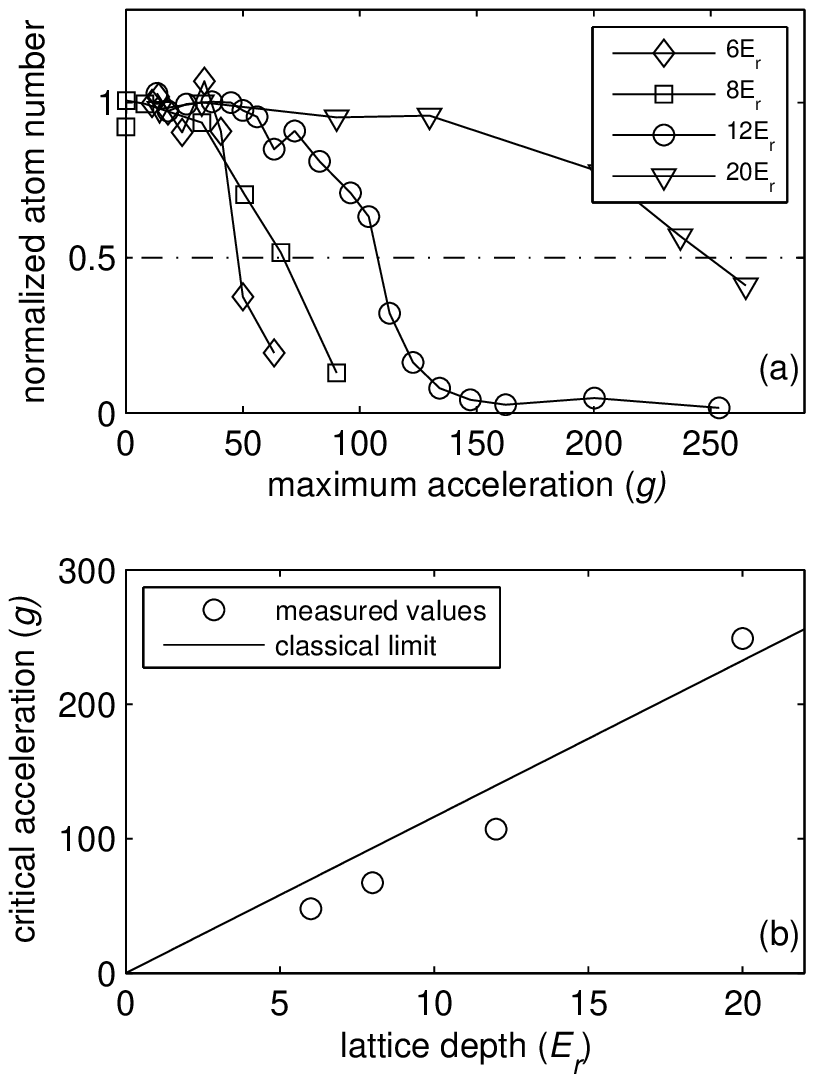}
\caption{ \textsc{Critical acceleration in lattice.} \textbf{(a)}
For several round-trip transports with varying maximum acceleration
$a$ and lattice depth (see legend), the number of remaining atoms
after transport is shown. As the maximum acceleration exceeds a
critical value, the number of atoms starts to drop significantly. We
define a critical acceleration as the maximum acceleration for
transports in which 50$\%$ of the atoms still reach their final
destination. This critical acceleration is shown as a function of
the lattice depth in \textbf{(b)}. The experimentally determined
values are compared with the limit expected from classical
considerations: $a_{crit} =  U_{\textrm{\scriptsize latt}}k / m$. }
\label{fig:amax}
\end{figure}

In Figure \ref{fig:becpics} we have studied the transport of a BEC,
which is especially sensitive to heating and instabilities. It is
important to determine, whether the atoms are still Bose-condensed
after the transport and what their temperature is afterwards. Figure
\ref{fig:becpics} shows momentum-distributions for various transport
distances $D$, which were obtained after adiabatically reloading the
atoms into the magnetic trap by ramping down the lattice and
subsequent time-of-flight measurements.

Before discussing these results, we point out that loading the BEC
adiabatically into the stationary optical lattice is already
critical. We observe a strong dependence of the condensate fraction
on the lattice depth. For too low lattice depths most atoms fall out
 of the lattice trap due to the gravitational field. For too high lattice
depths all atoms are trapped but the condensate fraction is very
small. One explanation for this is that high lattice depths lead to
the regime of 2D pancake shaped condensates where tunnelling between
adjacent lattice sites is suppressed. Relative dephasing of the
pancake shaped condensates will then reduce the condensate fraction
after release from the lattice. We obtain the best loading results
for a 11$\,E_r$ deep trap, where we lose about 65$\%$ of the atoms,
but maximize the condensate fraction. Because high lattice
intensities are detrimental for the BEC, we readjust the power of
the lattice during transport, such that the intensity is kept
constant over the transport range. The adjustments are based on the
calculated axial intensity distribution of the Bessel beam. In this
way we reach transport distances for BEC of 10$\,$cm. We believe,
that more sophisticated fine tuning of the power adjustments should
increase the transport length considerably. After transport
distances of $D$ = 18$\,$cm (36$\,$cm round trip) the atomic cloud
is thermal. Its momentum spread, however, is merely 0.3$\,\hbar k$,
which corresponds to a temperature of 30$\,$nK. Additionally, we
want to point out, that the loss of atoms due to the transport is
negligible ($<$10$\%$) compared to the loss through loading and
simply holding in such a low lattice potential ($\approx 65\%$).


An outstanding feature of the lattice transport scheme is the
precise positioning of the atomic cloud. Aside from uncontrolled
phase shifts due to residual mechanical noise, such as vibrating
optical components, we have perfect control over the relative phase
of the lattice lasers with our RF / AOM setup. This would in
principle result in an arbitrary accuracy in positioning the optical
lattice. We have experimentally investigated the positioning
capabilities in our setup. For this we measured in many runs the
position of the atomic cloud in the lattice after it had undergone a
return trip with a transport distance of $D = $ 10$\,$cm. The
position jitter, i.e. the standard deviation from the mean position,
was slightly below 1$\,\mu$m. For comparison, we obtain very similar
values for the position jitter when investigating BECs in the
lattice before transport. Hence the position jitter introduced
through the transport scheme is negligible.

 Another important property of the lattice transport
scheme is its high speed. For example, for a transport over 20$\,$cm
(40$\,$cm round trip) with negligible loss, a total transport time
of 200$\,$ms turns out to be sufficient. This is more than an order
of magnitude faster than in the MIT experiment \cite{Gus02}, where
an optical tweezer was mechanically relocated. The reason for this
speed up as compared to the optical tweezer is mainly the much
higher axial trapping frequency of the lattice  and the
non-mechanical setup.

In order to determine experimentally the lower limit of
transportation time we have investigated round-trip transports ($D
=$ 5$\,$mm), where we have varied the maximum acceleration and the
lattice depth (Figure \ref{fig:amax}(a)). The number of atoms, which
still remain in the lattice after transport, is measured. As soon as
the maximum acceleration exceeds a critical value, the number of
atoms starts to drop. For a given lattice depth, we define a
critical acceleration $a_{crit}$ as the maximum acceleration of the
particular transport where 50$\%$ of the atoms still reach their
final destination. Figure \ref{fig:amax}(b) shows the critical
acceleration $a_{\textrm{\scriptsize crit}}$ as a function of
lattice depth.
 The upper bound on acceleration observed here can be understood from classical
 considerations. In our lattice the maximum confining force along
 the axial direction is given by $U_{latt} k$, where $k$
 is the wave vector of the light field.
Thus in order to keep an atom bound to the lattice, we require the
acceleration $a$ to be small enough such that
\begin{equation}
 m a < U_{\textrm{\scriptsize latt}}k.
\end{equation}
Our data in Figure \ref{fig:amax} are in good agreement with this
limit \footnote{In the weak lattice regime ($U_{\textrm{\scriptsize
latt}} \ll 10\,E_{r}$) transport losses would be dominated by Landau
Zener tunnelling, see e.g. \cite{Pei97,Den02}.}.

There is in principle also a lower bound on the acceleration, which
is due to instabilities exhibited by BECs with repulsive
interactions loaded into periodic potentials
\cite{Cho99,Wu03,Fal04,Cri04}. Due to the fact that these
instabilities mainly occur at the edge of the Brillouin zones, the
time spent in this critical momentum range should be kept small. For
our lattice parameters nearly half of the Brillouin zone is an
unstable region, where the lifetime of the BEC is only on the order
of 10$\,$ms \cite{Fal04}. Thus we tend to sweep through the
Brillouin zone in much less than $\Delta t =$ 20$\,$ms, which
corresponds  to an acceleration of  $ a = \dot{v} \,  \gg \,
2v_{r}/\Delta t \simeq $ 0.6$\,$m/s$^{2}$. In this way BECs may be
transported without introducing too much heating through these
instabilities.

In contrast to acceleration, the transport velocity in our
experiment is only technically limited due to the finite AOM
bandwidth. As discussed before, the lattice is set in motion by
introducing a detuning between the two beams via AOMs
(equation~(\ref{vDelta})). For detunings exceeding the bandwidth of
the AOM, the diffraction efficiency of the modulator starts to drop
significantly. Consequently the lattice confinement vanishes, and
the atoms are lost. In our setup we can conveniently reach
velocities of up to $v=6\,$m/s $\approx$ 1100$\,v_{r}$,
corresponding to a typical AOM bandwidth of 15$\,$MHz. This upper
bound actually limits the transport time for long distance
transports ($D$ $>$ 5$\,$cm).

\begin{figure}
\centering
\includegraphics{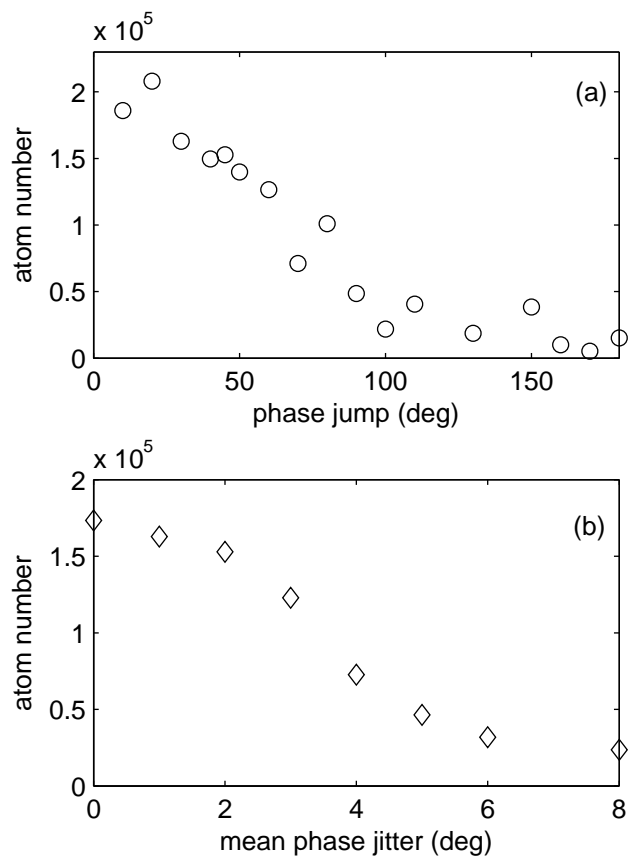}
\caption{ \textsc{Stability requirements for transport.} Sudden
phase jumps are introduced in the relative phase of the two
counterpropagating lattice laser beams. The corresponding abrupt
displacements of the optical lattice lead to heating and loss of the
atoms. We measure the number of atoms which remain in the lattice
after transport. \textbf{(a)}  Data obtained after a single relative
phase jump of variable magnitude. \textbf{(b)}  A phase jitter (200
positive Poissonian-distributed phase jumps with a variable mean
value) is introduced during transport. Mean values on the order of a
few degrees already lead to a severe loss of atoms.}
\label{fig:phase_paper}
\end{figure}

Finally we have investigated the importance of phase stability of
the optical lattice for the transport (see Figure
\ref{fig:phase_paper}). For this we purposely introduced sudden
phase jumps during transport to one of the lattice beams. The
timescale for the phase jumps, as given by AOM response time of
about 100$\,$ns, was much smaller than the inverse trapping
frequencies. The phase jumps lead to abrupt displacements of the
optical lattice, causing heating and loss of atoms. In Figure
\ref{fig:phase_paper}(a) the atomic losses due to a single phase
jump during transport are shown. Phase jumps of 60$^{\circ}$
typically induce a 50 \% loss of atoms. For continuous phase jitter
(see Figure 7(b)) the sensitivity is much larger.

\begin{figure}
\centering
\includegraphics{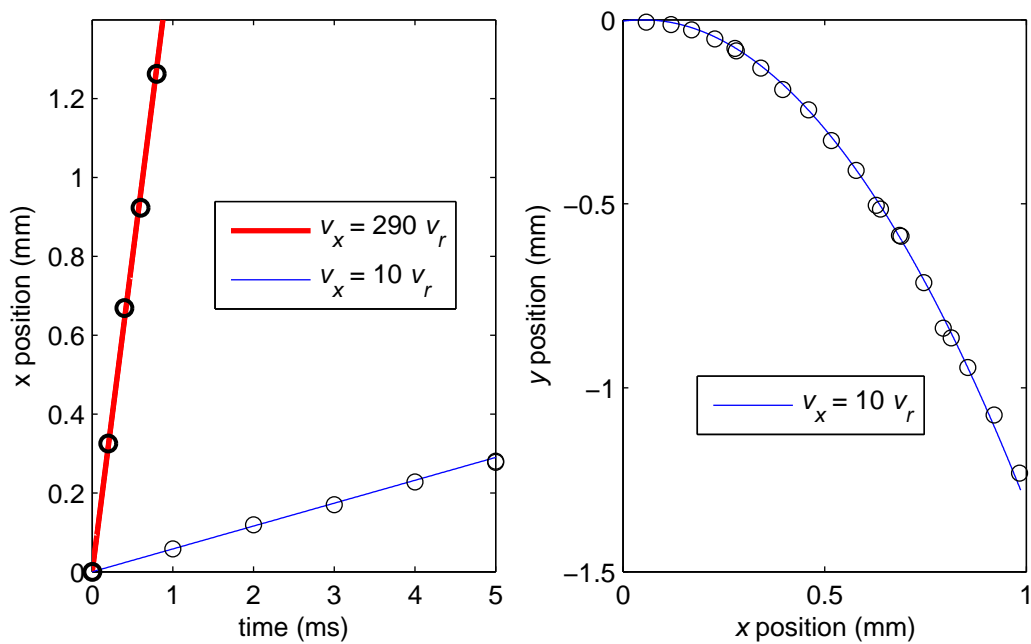}
\caption{ \textsc{Atom catapult.} After acceleration in
$x$-direction and subsequent release from the lattice, the position
of the atomic cloud is tracked as it flies ballistically through the
field of view of the CCD camera. Shown are two data sets where atoms
were accelerated to velocities of either $v_{\textrm{\scriptsize
x}}$ = 10$\,v_{\textrm{\scriptsize r}}$ or $v_{\textrm{\scriptsize
x}}$ = 290$\,v_{\textrm{\scriptsize r}}$. \textbf{(a)} The
horizontal position $x$ as a function of time. \textbf{(b)} For the
slower cloud ($v_{\textrm{\scriptsize x}}$ =
10$\,v_{\textrm{\scriptsize r}}$) a parabolic trajectory
$y=-g/2\cdot (x/v_{\textrm{\scriptsize x}})^{2}$ is observed as it
falls under the influence of gravity.} \label{fig:catapult}
\end{figure}

\section{Atom catapult} In addition to transport of ultracold atoms,
acceleration of atoms to precisely defined velocities is another
interesting application of the moving optical lattice.  For
instance, it could be used to study collisions of BECs with a very
high but well defined relative velocity, similar to the experiments
described in \cite{Tho04,Bug04}. As already shown above, we have
precise control to impart a well defined number of up to 1100 photon
recoils to the atoms. This corresponds to a large kinetic energy of
$k_B\,\times\, $200$\,$mK. At the same time the momentum spread of
the atoms is about 1/3 of a recoil (see Figure 5). To illustrate
this, we have performed two sets of experiments, where we accelerate
a cloud of atoms to velocities $v = 10\,v_{r}$ and $v = 290\,v_{r}
\approx$ 1.6$\,$m/s. After adiabatic release from the lattice, we
track their position in free flight (see Figure \ref{fig:catapult}).
Initially the atomic cloud is placed about 8$\,$cm away from the
position of the magnetic trap. It is then accelerated back towards
its original location. Before the atoms pass the camera's field of
vision, the lattice beams are turned off within about 5 ms, to allow
a ballistic flight of the cloud. Using absorption imaging the
position of the atomic cloud as a function of time is determined.
The slope of the straight lines in Figure \ref{fig:catapult}(a)
corresponds nicely to the expected velocity. However, due to a time
jitter problem, individual measurements are somewhat less precise
than one would expect\footnote{This is linked to the fact that our
clock for the system control is synchronized to the 50 Hz of the
power grid. Fluctuations of the line frequency lead to shot to shot
variations in the ballistic flight time of the atoms, which
translates into an apparent position jitter.}. For $v =
10$\,$v_{r}$, Figure~\ref{fig:catapult}(b) shows the trajectory of
the ballistic free fall of the atoms in gravity.

\section{Conclusion}
In conclusion, we have realized a long distance optical transport
for ultracold atoms, using a moveable standing wave dipole trap.
With the help of a diffraction-free Bessel beam, macroscopic
distances are covered for both BEC and ultracold thermal clouds. The
lattice transport features a fairly simple setup, as well as a fast
transport speed and high positional accuracy. Limitations are mainly
technical and leave large room for improvement. In addition to
transport, the lattice can also be used as an accelerator to impart
a large but well defined number of photon recoils to the atoms.

\section{Acknowledgements}
We want to thank U. Schwarz for helpful information on the
generation of Bessel beams and for lending us phase-gratings for
testing purposes. Furthermore we thank R. Grimm for discussions and
support. This work was supported by the Austrian Science fund (FWF)
within SFB 15 (project part 17) and the Tiroler Zukunftsstiftung.

\section*{Appendix: Transport Ramp}
We give here the analytic  expression for the lattice acceleration
$a(t)$ as a function of time $t$ which was implemented in our
experiments (see for example Figure 3). $a(t)$ is a smooth piecewise
defined cubic polynomial,
\begin{equation*}
  \! \! \! \! \! \! \! \! \! \! \! \! \! \! \! \! \! \! \! \! \! \! \! \! \! \! \! a(t) =\left\{
  \begin{array}{r
      @{\quad{\mbox{for}}\quad}
      l}
    \frac{D}{T^{2}}\;(-\frac{7040}{9}
(\frac{t}{T})^{3}+320 (\frac{t}{T})^{2}) & 0 < t \leq T/4 \\
     \frac{D}{T^{2}}\;(\frac{3200}{9}
(\frac{t}{T})^{3}-\frac{1600}{3}
(\frac{t}{T})^{2}+\frac{640}{3} \frac{t}{T}-\frac{160}{9}) & T/4 < t \leq 3T/4 \\
    \frac{D}{T^{2}}\;(-\frac{7040}{9}
(\frac{t}{T})^{3}+\frac{6080}{3} (\frac{t}{T})^{2}-\frac{5120}{3}
\frac{t}{T}
+\frac{4160}{9}) & 3T/4 < t \leq T \\
  \end{array}
  \right.
\end{equation*}
Here, $D$ is the distance over which the lattice is moved and $T$ is
the duration of the transport. From $a(t)$ both the velocity $v(t)$
and the location $x(t)$ may be derived via integration over time.
Our choice for the acceleration $a(t)$ features a very smooth
transport. The acceleration $a(t)$ and its derivative $\dot a(t)$
are zero at the beginning ($t = 0$) and at the end ($t = T$) of the
transport. At $t=T/4$ and $t=3T/4$ the absolute value of the
acceleration reaches a maximum.

\section*{References}

\input{bibliography}

\end{document}